\documentclass[letterpaper]{article} % DO NOT CHANGE THIS
\usepackage{aaai23}  % DO NOT CHANGE THIS
\usepackage{times}  % DO NOT CHANGE THIS
\usepackage{helvet}  % DO NOT CHANGE THIS
\usepackage{courier}  % DO NOT CHANGE THIS
\usepackage[hyphens]{url}  % DO NOT CHANGE THIS
\usepackage{graphicx} % DO NOT CHANGE THIS
\usepackage{natbib}  % DO NOT CHANGE THIS AND DO NOT ADD ANY OPTIONS TO IT
\usepackage{caption} % DO NOT CHANGE THIS AND DO NOT ADD ANY OPTIONS TO IT
\usepackage{algorithm}
\usepackage{algorithmic}

\usepackage{newfloat}
\usepackage{listings}
\DeclareCaptionStyle{ruled}{labelfont=normalfont,labelsep=colon,strut=off} % DO NOT CHANGE THIS
\floatstyle{ruled}
\newfloat{listing}{tb}{lst}{}
\floatname{listing}{Listing}
\title{edBB-Demo: Biometrics and Behavior Analysis for Online Educational Platforms}
\author{
    Roberto Daza,
    Aythami Morales,
    Ruben Tolosana,
    Luis F. Gomez,
    Julian Fierrez,
    Javier Ortega-Garcia
}
\affiliations{
    School of Engineering, Autonomous University of Madrid\\
    \{roberto.daza, aythami.morales, ruben.tolosana, luisf.gomez, julian.fierrez, javier.ortega\}@uam.es

}

\usepackage{bibentry}
\begin{document}

\maketitle

\begin{abstract}
We present edBB-Demo, a demonstrator of an AI-powered research platform for student monitoring in remote education. The edBB platform aims to study the challenges associated to user recognition and behavior understanding in digital platforms. This platform has been developed for data collection, acquiring signals from a variety of sensors including keyboard, mouse, webcam, microphone, smartwatch, and an Electroencephalography band. The information captured from the sensors during the student sessions is modelled in a multimodal learning framework. The demonstrator includes: \textit{i)} Biometric user authentication in an unsupervised environment; \textit{ii)} Human action recognition based on remote video analysis; \textit{iii)} Heart rate estimation from webcam video; and \textit{iv)} Attention level estimation from facial expression analysis.    
\end{abstract}

\section{Introduction}

\noindent Our society is moving from the physical to the digital world. One example is in virtual education. The e-learning industry has grown over $5$\% per year since the last decade \cite{WEFsurvey2016}. The interest in online educational platforms has been motivated by the need of a continuous learning process during the entire professional life and the worldwide access to high-quality content. 

Virtual education avoids the traditional spatial constraints of a class, and permits a higher number of people to access the same training contents at the same time. In addition, the flexibility of virtual education gives students the possibility to connect to the teaching platform at any time and any place, compared with traditional education that establishes strict schedules and mandatory physical attendance. As a result of this flexibility and the worldwide access to the internet, students have a chance of studying independently of their location and schedule. However, this does not guarantee a more effective or faster learning process compared to traditional teaching. Additionally, some problems may appear derived from the dependence of a functional Internet connection, a device, or the own teaching platform. The learning process can also be challenging for students due to the distant relation between them and the teachers. As a result, virtual education presents some advantages but also important drawbacks. Among the most important ones we can highlight the detection of fraud or cheating in a unsupervised environment and the lack of face-to-face interaction between the lecturers and students in asynchronous courses (e.g., pre-recorded content). The feedback generated during the face-to-face interaction provides useful information to better adapt the contents and guide the learning process during the class.

To mitigate these main drawbacks of virtual education, approaches based on biometrics and behavior analysis can be a good solution. The recent advances in machine learning and digital behavior understanding allow to classify and monitor users by their physiological and behavioral characteristics \cite{jain201650}. This can be done in a transparent and user-friendly way during the whole session of the student. Behavior understanding can be used to adapt dynamically the environment and teaching content or improve the educational resources with a further analysis of the e-learning sessions. 

\begin{figure}[t]
\centering
\includegraphics[width=1\columnwidth]{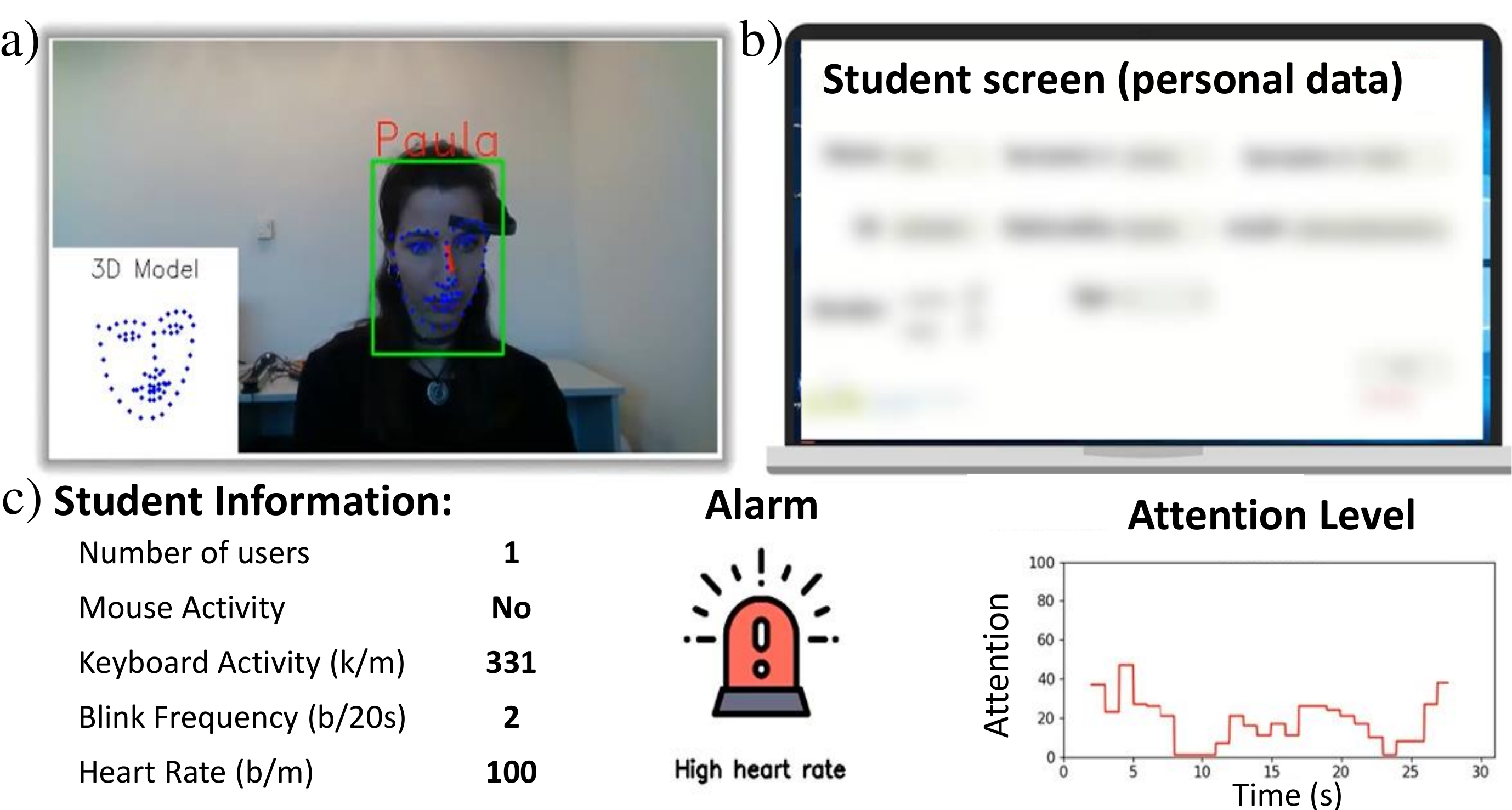} % Reduce the figure size so that it is slightly narrower than the column.
\caption{edBB-Demo: a) front-end with the webcam video, b) student screen, and c) biometric and behavior modules.}
\label{demo}
\end{figure}

\begin{figure*}[t]
\centering
\includegraphics[width=1\textwidth]{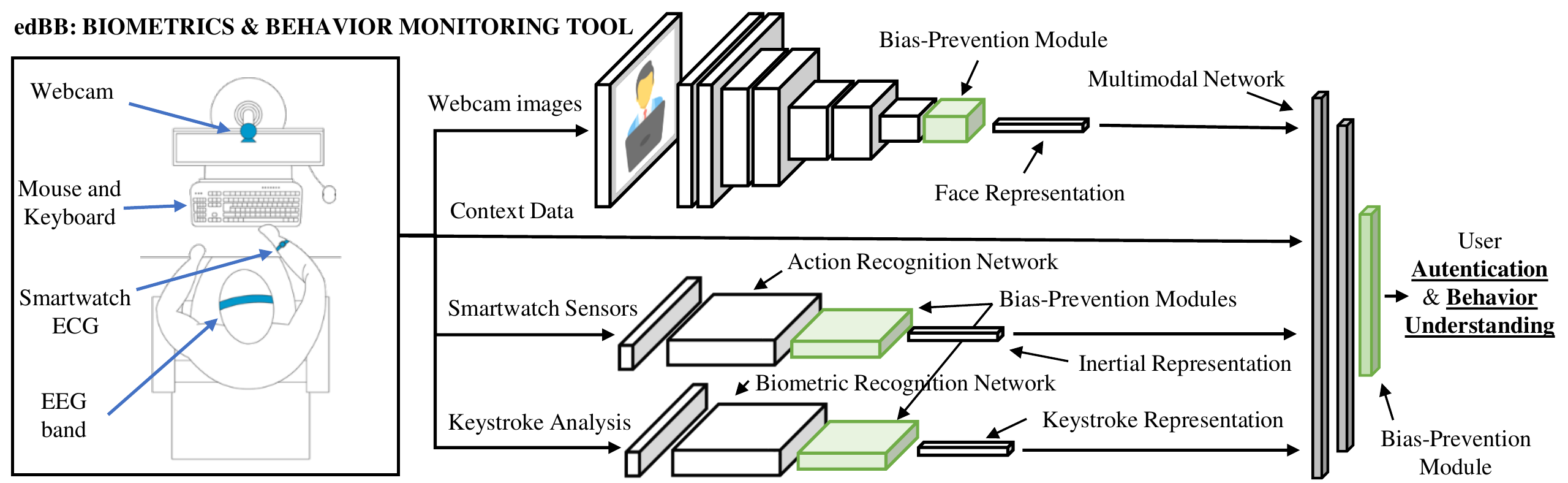} % Reduce the figure size so that it is slightly narrower than the column.
\caption{Block diagram of the edBB-Demo back-end. Information captured by the edBB platform and its integration into the multimodal learning framework.}
\label{block_diagram}
\end{figure*}

The edBB-Demo aims to explore how AI technologies can improve learning outcomes and student engagement by incorporating biometrics and behavioral information into the e-learning sessions. However, the usage of these technologies arise important privacy and security concerns that need to be discussed and addressed \cite{jain2021biometrics}. The edBB-Demo and the databases \cite{hernandez2019edbb} generated with the platform will provide useful resources to study the bias and trustworthiness elements of these biometrics-improved e-learning platforms.

\section{edBB-Demo: AI Technologies and Challenges}

Fig. \ref{demo} shows the main functionalities of the edBB-Demo. The figure shows the front-end of the demonstrator during a student session\footnote{See the full video at: https://youtu.be/JbcL2N4YcDM}. The demo integrates the outputs of the different biometrics and behavior models into the front-end. It also includes an alarm signal that is activated when pre-defined events are detected (e.g., high heart rate). 

edBB-Demo incorporates a variety of technologies for biometrics and behavior analysis (see Fig. \ref{block_diagram}). These technologies work as independent modules (e.g., heart rate estimation) or in combination (e.g., face and keystroke authentication). edBB-Demo is a demonstrator of both \textit{i)} the main advances in AI applications for student monitoring, and \textit{ii)} the challenges associated to biometrics modelling and behavior understanding in digital platforms. We highlight next the key modules included in edBB-Demo:    

    \textbf{Biometric User Recognition:} the platform includes a multimodal biometric system based on the fusion of face and keystroke dynamics. The performance of face recognition models has outperformed human capabilities in the last decade \cite{deng2019arcface}. edBB-Demo includes a ResNet-50 model trained with more than $64,000$ identities from heterogeneous ethnics \cite{serna2022sensitive}. However, face recognition technology is vulnerable to physical and digital attacks \cite{rathgeb2022handbook}, and the unsupervised nature of e-learning sessions increases that vulnerability. Keystroke authentication is used in combination with face as a two-factor authentication (2FA). edBB includes an LSTM keystroke recognition model trained with more than $4$M keystroke samples \cite{acien2021typenet}.      
    
    \textbf{Action Recognition:} head pose estimation from 2D face images is used in combination with inertial sensors from external devices (e.g., smartwatches) to model the user behavior during  online sessions. Modelling the user behavior is a challenging task in machine learning applications. The head pose is estimated using a Convolutional Neural Network based on \cite{berral2021realheponet}. In addition, the platform captures the screen of the student's computer to detect content not allowed during the session (e.g., in exams).
    
    \textbf{Heart Rate Estimation:} the heart rate is estimated using remote Photoplethysmography (rPPG) from face videos \cite{hernandez2020heart}. The heart rate signal from the smartwatch is used as ground-truth data to train and evaluate the models. Heart rate can be used as a biomarker of the emotional state, providing useful insights about the stress level of the student.  
    
    \textbf{Attention Level Estimation:} the attention is modelled using the faces gestures (e.g., blink rate) and user actions (e.g., head pose). The ground-truth of the attention level is obtained with an EEG band. The attention level provides insights about the cognitive load of the student during the online sessions. edBB incorporates an attention level estimation from the webcam video \cite{daza2020mebal, daza2021alebk}. 
    
    \textbf{Discrimination-aware Learning:} privacy-preserving and discrimination-aware technologies are important concerns for our society. edBB-Demo incorporates an agnostic learning method \cite{morales2020sensitivenets} to reduce the impact of sensitive features in the automatic-decision process (e.g., gender or ethnicity when detecting cheaters). The aim of the bias prevention modules (see Fig. \ref{block_diagram}) is to remove the demographic patterns (e.g., gender and ethnicity) from the feature space of the trained models. 
    
\section{Conclusions}

The edBB-Demo presents some of the most important advances made during the last decade in remote biometrics and behavioral understanding. These technologies are combined into a multimodal learning framework to explore the present and future capabilities of virtual education platforms. Together with the advantages, this demonstrator is a path to discuss the security and privacy concerns associated to these technologies. The edBB-Demo includes the edBB-db, a multimodal database with sessions of $60$ students (more than $20$ minutes per session). This database comprises information from all sensors of the platform acquired in a realistic e-learning environment (i.e., a psychotechnical exam).  

\section{Acknowledgments}

Support by projects: BBforTAI (PID2021-127641OB-I00 MICINN/FEDER), TRESPASS-ETN (MSCA-ITN-2019-860813), BIBECA (RTI2018-101248-B-I00 MINECO/FEDER) and by BIOPROCTORING. Roberto Daza is supported by a FPI fellowship from the Spanish MINECO/FEDER.

\bibliography{aaai23}

\end{document}